\documentclass[10pt,twocolumn]{article}
\usepackage{times}
\usepackage{latex8}
\usepackage[dvips]{graphicx}
\usepackage{subfigure}
\usepackage{algorithm}
\usepackage{algorithmic}

\begin{document}

\title{Ad-hoc Limited Scale-Free Models for Unstructured Peer-to-Peer
Networks}

\author{Hasan Guclu\\
Center for Nonlinear Studies\\
Los Alamos National Laboratory\\
Los Alamos, NM 87545. guclu@lanl.gov\\
\and
Durgesh Kumari and Murat Yuksel\\
Computer Science and Engineering Department\\
University of Nevada - Reno, Reno, NV 89557.\\
durgesh.rani@gmail.com yuksem@cse.unr.edu\\
}

\maketitle \thispagestyle{empty}

\begin{abstract}
Several protocol efficiency metrics (e.g., scalability, search
success rate, routing reachability and stability) depend on the
capability of preserving structure even over the churn caused by the
ad-hoc nodes joining or leaving the network. Preserving the
structure becomes more prohibitive due to the distributed and
potentially uncooperative nature of such networks, as in the
peer-to-peer (P2P) networks. Thus, most practical solutions involve
unstructured approaches while attempting to maintain the structure
at various levels of protocol stack. The primary focus of this paper
is to investigate construction and maintenance of scale-free
topologies in a distributed manner without requiring global topology
information at the time when nodes join or leave. We consider the
\emph{uncooperative behavior} of peers by limiting the number of
neighbors to a pre-defined hard cutoff value (i.e., no peer is a
major hub), and the \emph{ad-hoc behavior} of peers by rewiring the
neighbors of nodes leaving the network. We also investigate the
effect of these hard cutoffs and rewiring of ad-hoc nodes on the P2P
search efficiency.
\end{abstract}


\Section{Introduction}

Stability and scalability of highly dynamic networks mainly depends
on the capability of preserving structure even over the churn caused
by the ad-hoc nodes joining or leaving the network. Several protocol
efficiency metrics (e.g., search success rate, routing reachability
rate) depend on this capability. Preserving the structure becomes
more prohibitive due to the distributed and potentially
uncooperative nature of such networks, as in the peer-to-peer (P2P)
networks. Thus, most practical solutions involve unstructured
approaches while attempting to maintain the structure at various
levels of protocol stack.

In decentralized P2P networks, the overlay topology (or connectivity
graph) among peers is a crucial component in addition to the
peer/data organization and search. Topological characteristics have
profound impact on the efficiency of search on P2P networks as well
as other networks. It has been well-known that search on small-world
topologies can be as efficient as $O(\ln N)$ \cite{kleinberg-2000},
and this phenomenon has recently been studied on P2P networks
\cite{ZGG02,MSZ-2005,HLY-2006,IRF04}. The best search efficiency in
realistic networks can be achieved when the topology is scale-free
(power-law), which offers search efficiencies like $O(\ln \ln N)$.
Key limitation of scale-free topologies is the high load (i.e., high
degree) on very few number of hub nodes. In a typical unstructured
P2P network, peers are not willing to maintain high degrees/loads as
they may not want to store large number of entries for construction
of the overlay topology. So, to achieve fairness and practicality
among all peers, \emph{hard} cutoffs on the number of entries are
imposed by the individual peers, which makes the overall network a
``limited'' one. Effect of such hard cutoffs on search efficiency
can be significant \cite{guclu-2007}.

Due to the uncooperative nature of peers in a P2P network, protocols
cannot completely rely on methods working with full cooperation of
peers. For example, peers may not want to store large number of
entries for construction of the overlay topology, i.e., connectivity
graph. Even though characteristics of the overlay topology is
crucial in determining the efficiency of the network, peers
typically do not want to take the burden of storing excessive amount
of control information for others in the network, thereby imposing
\emph{hard cutoffs} on the amount of control information to be
stored. Yet another key issue is the construction of scale-free
overlay topologies without global information. There are several
techniques to generate a scale-free topology \cite{bara99,albert00},
by using \emph{global} information about the current network when a
node joins or leaves. Such global methods are not practical in P2P
networks, and \emph{local} heuristics in generating such scale-free
overlay topologies must be employed. In other words, there must be
local and simple operations when peers are joining or leaving the
P2P overlay, and also causing a minimal inefficiency to the search
mechanisms to be run on the network.

The primary focus of this paper is to investigate construction and
maintenance of scale-free topologies in a distributed manner without
requiring global topology information at the time when nodes join or
leave. We consider the \emph{uncooperative behavior} of peers by
limiting the number of neighbors to a pre-defined hard cutoff value,
and the \emph{ad-hoc behavior} of peers by rewiring the neighbors of
nodes leaving the network. We also investigate the effect of these
hard cutoffs and the rewiring of ad-hoc nodes on the P2P search
efficiency.

The rest of the paper is organized as follows: First, we provide
motivation for this work, outline the key parameters to be
considered, and briefly state the major contributions and findings
of the work. Then, we survey the previous work on P2P networks in
Section~\ref{sec:survey}. In
Section~\ref{sec:scale-free-topologies}, we survey the previous work
on scale-free topology generation and briefly cover the importance
of \emph{cutoff} in the scale-free network and \emph{Preferential
Attachment (PA) with Hard Cutoffs}. We introduce our topology
generation techniques using local heuristics and briefly describe
the algorithm showing join and leave process of a node in the
growing network, in Section~\ref{sec:local-heuristics}. In
Section~\ref{sec:simulations}, we present our simulation results of
degree distribution of the nodes. We also discuss the efficiency of
three search algorithms {i.e Flooding (FL), Normalized Flooding (NF)
and Random Walk (RW) on the topology generated by our simulations.
We conclude by summarizing our current work and outlining the
directions for the future work in Section~\ref{sec:summary}.

\SubSection{Contributions and Major Results}

%
%
%
%

Our work uncovers the relationship between the ad-hoc behavior of
peers (i.e., how frequent they join/leave) and the efficiency of
search over an overlay topology where each peer can (or is willing
to) store a maximum number of links to other peers. In our model, we
parameterize (i) the ad-hoc behavior of nodes by the probability
that a node leaves $\mu$, (ii) the amount of local information to be
used at the time of \emph{join} by knowledge radius from the point
the node attempts to join, $\tau_j$ (i.e., the node knows about the
local topology covering $\tau_j$ hops away from the point the node
attempts to join the network), and (iii) the amount of local
information to be used at the time of \emph{leave} by knowledge
radius from the location of the leaving node, $\tau_l$ (i.e., each
neighbor of the leaving node knows about the local topology covering
$\tau_l$ hops away from itself). We also define the maximum number
of links to be stored by peers as the \emph{hard cutoff}, $k_c$, for
the degree of a peer in the network as compared to \emph{natural
cutoff} which occurs due to finite-size effects. Our contributions
include:
\begin{itemize}
\item{\emph{Guidelines for generating scale-free topologies over ad-hoc nodes:}}
We introduce a generic model that can assign availability of
different amount of local topology information at the times when a
node joins or leaves. Our model provides a way of tuning ad-hocness
of the network and studying how to balance state information for
nodes joining or leaving.

\item{\emph{Search efficiency on ad-hoc limited scale-free
topologies:}} Through extensive simulations, we studied efficiency
of Flooding (FL), Normalized Flooding (NF), and Random Walk (RW) on
the topologies generated by our model with different $\mu$,
$\tau_j$, $\tau_l$, and $k_c$ values.

\item{\emph{Rewiring methodologies for designing peer leave algorithms for
unstructured P2P networks:}} Our study yielded several guidelines
for peers leaving an unstructured P2P network, so that the search
performance of the overall overlay topology remains high.
\end{itemize}

Our study revealed several interesting issues. We found that having
more global information about the topology at the time of leave is
significantly more helpful than having it at the time of join. We
show that the degree distribution can be kept scale-free and the
search efficiency can be kept very high by simply keeping $\tau_l$
at reasonably high values, e.g., 2-3.

\Section{Related Work} \label{sec:survey}
Previous work on P2P network protocols can be classified into
\emph{centralized} and \emph{decentralized} ones. As centralized P2P
protocols (e.g. Napster \cite{napster}) proved to be unscalable, the
majority of the P2P research has focused on decentralized schemes.
The decentralized P2P schemes can be further classified into
sub-categories: \emph{structured}, \emph{unstructured}, and
\emph{hybrid}.

In the structured P2P networks, data/file content of peers is
organized based on a keying mechanism that can work in a distributed
manner, e.g. Distributed Hash Tables (DHTs) \cite{chord01}. The
keying mechanism typically maps the peers (or their content) to a
logical search space, which is then leveraged for performing
efficient searches.
In contrast to the structured schemes, unstructured P2P networks do
not include a strict organization of peers or their content. Since
there is no particular keying or organization of the content, the
search techniques are typically based on flooding. Thus, the
searches may take very long time for rare items, though popular
items can be found very fast due to possible leveraging of locality
of reference \cite{SMZ03} and caching/replication \cite{CoSh-2002}.

%
%

The main focus of the research on unstructured P2P networks has been
the tradeoff between state complexity of peers (i.e., number of
records needed to be stored at each peer) and flooding-based search
efficiency. The minimal state each peer has to maintain is the
\emph{list of neighbor peers}, which construct the overlay topology.
Optionally, peers can maintain \emph{forwarding table}s (also
referred as routing tables in the literature) for data items in
addition to the list of neighbor peers. Thus, we can classify
unstructured P2P networks into two based on the type(s) of state
peers maintain: (i) \emph{per-data} unstructured P2P networks (i.e.,
peers maintain both the list of neighbor peers and the per-data
forwarding table), and (ii) \emph{non-per-data} unstructured P2P
networks (i.e., peers maintain only the list of neighbor peers).

Non-per-data schemes are mainly Gnutella-like schemes
\cite{gnutella}, where search is performed by means of flooding
query packets. Search performance over such P2P networks has been
studied in various contexts, which includes pure random walks
\cite{GMS04}, probabilistic flooding techniques \cite{KXZ05,GMS05},
and systematic filtering techniques \cite{Xu03}.

Per-data schemes (e.g. Freenet \cite{freenet}) can achieve better
search performances than non-per-data schemes, though they impose
additional storage requirements to peers. By making the peers
maintain a number of $<$key,pointer$>$ entries peers direct the
search queries to more appropriate neighbors, where ``key'' is an
identifier for the data item being searched and the ``pointer'' is
the next-best neighbor to reach that data item. This capability
allows peers to leverage associativity characteristics of search
queries \cite{CFK03}. Studies ranged from grouping peers of similar
interests (i.e., peer associativity) \cite{IRF04,CFK03} to
exploiting locality in search queries (i.e., query associativity)
\cite{CRBLS03,SMZ03}. Our work is applicable to both per-data and
non-per-data unstructured P2P networks, since we focus on the
interactions between search efficiency and topological
characteristics.

Previous study on node isolation caused by churn in unstructured P2P
networks introduced a general model of resilience \cite{YWLL07}. In
this study, joining and rewiring processes were based on age-biased
neighbor selection, where a formal analysis included two age-biased
techniques of neighbor selection. In maximal age-selection approach,
the joining node selects uniformly randomly $m$ alive nodes from the
network and connects to the one with maximal age. It follows the
same process when a dead link is detected. However, in age-biased
random walk selection approach, the probability of a node to be
selected by another peer is proportional to its current age. Another
study introduced self-organizing super peer network architecture
\cite{GES07}, where super peers maintain the cache with pointers to
files that are recently requested and on the other hand client peers
dynamically select super peers offering best search results.

\Section{Scale-Free Network Topologies}
\label{sec:scale-free-topologies}

Recent research shows that many natural and artificial systems such
as the Internet \cite{faloutsos99}, World Wide Web
\cite{albert00_2}, scientific collaboration network \cite{bara02},
and e-mail network \cite{ebel02} have power-law degree
(connectivity) distributions. These systems are commonly known as
power-law or scale-free networks since their degree distributions
are free of scale (i.e., not a function of the number of network
nodes $N$) and follow power-law distributions over many orders of
magnitude. This phenomenon has been represented by the probability
of having nodes with $k$ degrees as $P(k)$$\sim$$k^{-\gamma}$ where
$\gamma$ is usually between $2$ and $3$ \cite{bara99}. Scale-free
networks have many interesting properties such as high tolerance to
random errors and attacks (yet low tolerance to attacks targeted to
hubs) \cite{alb00}, high synchronizability
\cite{guclu05,guclu07_chaos,korniss06}, and resistance to congestion
\cite{toroczkai04}.

The origin of the scale-free behavior can be traced back to two
mechanisms that are present in many systems, and have a strong
impact on the final topology \cite{bara99}. First, networks are
developed by the addition of new nodes that are connected to those
already present in the system. This mechanism signifies continuous
expansion in real networks. Second, there is a higher probability
that a new node is linked to a node that already has a large number
of connections. These two features led to the formulation of a
growing network model first proposed by Barab\'asi and Albert that
generates a scale-free network for which $P(k)$ follows a power law
with $\gamma$$=$$3$. This model is known as \textit{preferential
attachment} (PA or rich-gets-richer mechanism) and the resulting
network is called \textit{Barab\'asi-Albert} network
\cite{bara99,albert00}.

In this study, we use a simple version of the PA model
\cite{bara99}. The model evolves by one node at a time and this new
node is connected to $m$ (number of stubs) different existing nodes
with probability proportional to their degrees, i.e.,
$P_i$$=$$k_i/\sum_j k_j$ where $k_i$ is the degree of the
node $i$. The average degree per node in the resulting network is
$2m$ and the minimum degree is $m$.

Scale-free networks are very robust against random failures and
attacks since the probability to hit the hub nodes (few nodes with
very large degree) is very small and attacking the low-degree
satellite nodes does not harm the network. On the other hand,
deliberate attacks targeted to hubs through which most of the
traffic go can easily shatter the network and severely damage the
overall communication in the network. For the same reason the
Internet is called ``robust yet fragile'' \cite{doyle05} or
``Achilles' heel'' \cite{alb00, alb00_2}.

Scale-free networks also have \textit{small-world} \cite{watts98}
properties. In small-world networks the diameter, or the mean hop
distance between the nodes scales with the system size (or the
number of network nodes) $N$ logarithmically, i.e., $d$$\sim$$\ln N$.
The scale-free networks with $2$$<$$\gamma$$<$$3$ have a much
smaller diameter and can be named \textit{ultra-small} networks
\cite{cohen03}, behaving as $d$$\sim$$\ln{\ln N}$. When
$\gamma$$=$$3$ and $m$$\geq$$2$, $d$ behaves as
$d$$\sim$$\ln N/\ln\ln N$. However, when $m$$=$$1$ and
$\gamma$$=$$3$ the Barab\'asi-Albert model turns into a tree and
$d$$\sim$$\ln N$ is obtained. Also when $\gamma$$>$$3$, the diameter
behaves logarithmically as $d$$\sim$$\ln N$.
Since the
speed/efficiency of search algorithms strongly depend on the average
shortest path, scale-free networks have much better performance in
search than other random networks.


\SubSection{The Cutoff}

One of the important characteristics of scale-free networks is the
natural cutoff on the degree (or the maximum degree) due to
finite-size effects. Natural cutoff can be defined as
\cite{doro02_2} the value of the degree above which one expects to
find at most one vertex, i.e.,
\begin{equation}
N \int_{k_{nc}}^{\infty} P(k)dk \sim 1
\;.
\label{eq3}
\end{equation}
By using the degree distribution for the scale-free network
and the exact form of probability distribution (i.e.,
$P(k)$$=$$(\gamma-1)m^{\gamma-1}/k^{\gamma}$), one obtains
\begin{equation}
k_{nc}(N) \sim mN^{1/(\gamma-1)},
\label{eq4}
\end{equation}
which is known as the \textit{natural} cutoff of the network. The
scaling of the natural cutoff can also be calculated by using the
extreme-value theory \cite{boguna04}. For the scale-free networks
generated by PA model ($\gamma$$=$$3$) the
natural cutoff becomes
\begin{equation}
k_{nc}(N) \sim m\sqrt{N}.
\label{eq5}
\end{equation}

\renewcommand{\baselinestretch}{1}
\begin{algorithm}
\caption{Network growth using paramaterized join and
leave processes} \label{alg:growth}

\emph{//Global Variables and Functions}\\
\textbf{$m$} - \emph{minimum degree}\\
\textbf{$\mu$} - \emph{probability of  a node to leave the network}\\
\textbf{$N$} - \emph{the maximum node ID of the existing network
(the minimum node ID is 0)}\\
\textbf{$G$} - \emph{graph of the existing network of $M$ links and $N$ nodes}\\
\textbf{PreferentialAttachment($G_1$, $G_2$)} - \emph{a function
that performs Preferential Attachment to $G_1$ by using the nodes
in $G_2$, returns the number of successful new links}\\

\emph{// Join process of node i}\\
void \textbf{Join}(i, $\tau_j$)
\begin{algorithmic}[1]
\STATE N++ \STATE $numoflinks \leftarrow 0$

\WHILE {$numoflinks < m$}

\STATE $N_{rand} \leftarrow$ Randomize(1,$N$)
\COMMENT{Pick a random
node from the existing network}

\STATE $myG \leftarrow get\_subgraph(N_{rand} ,\tau_j)$
\COMMENT{Get
the subgraph including neighbor nodes of $N_{rand}$ up to $\tau_j$
hops away}

\STATE $numoflinks$ += PreferentialAttachment($G$, $myG$)

\ENDWHILE
\end{algorithmic}

\vspace{2mm} \emph{//Leave process of node i}\\
void \textbf{Leave}(i, $\tau_l$)
\begin{algorithmic}[1]
\STATE $myG \leftarrow get\_subgraph(N_{rand}, \tau_l)$ \COMMENT{Get
the subgraph including neighbor nodes of $N_{rand}$ up to $\tau_l$
hops away}

\STATE $remove(N_{rand})$ \COMMENT{Delete $N_{rand}$ from the
existing network} \STATE $N=N-1$ \STATE
PreferentialAttachment($G$,$myG$)
\end{algorithmic}

\vspace{2mm}\emph{// Growth process of a network with $N_{target}$
nodes, parameterized with $\tau_j$ and $\tau_l$}\\
void \textbf{Grow}($N_{target}$, $\tau_j$, $\tau_l$)
\begin{algorithmic}[1]
\FOR {i=$m$+1; i$<$$N_{target}$; i++}

\STATE Join(N,$\tau_j$)

\STATE $num \leftarrow$ Random(0,1)
\IF {$N == N_{target}$} \STATE break; \ENDIF

\IF {$num < \mu$}

\STATE $N_{del} \leftarrow$ Randomize(1,N)

\STATE Leave($N_{del}$, $\tau_l$)

\ENDIF \ENDFOR
\end{algorithmic}
\end{algorithm}

\SubSection{Preferential Attachment with Hard Cutoffs}

The natural cutoff may not be always attainable for most of the
scale-free networks due to technical reasons. One main reason is
that the network might have limitations on the number of links the
nodes can have. This is especially important for P2P networks in
which nodes can not possibly connect many other nodes. This requires
putting an artificial or \textit{hard} cutoff $k_c$ to the number of
links one node might have.

In order to implement the hard cutoff in PA, we simply did not allow
nodes to have links more than a fixed hard cutoff value during the
attachment process. This modified method generates a scale-free
network in which there are many nodes with degree fixed to hard
cutoff instead of a few very high degree hubs and the degree
distribution still decays in a power law fashion. The degree
distribution of PA model with cutoff is slightly different than that
of PA without a cutoff in terms of exponent and an accumulation of
nodes with degree equal to hard cutoff. PA model, in its original
form, has a degree distribution exponent $\gamma$$=$$3$ for very
large networks. However, when a hard cutoff is imposed it is
observed that the absolute value of degree distribution exponent
decreases \cite{guclu-2007}.

One can use the master-equation \cite{KRAPIVSKY01} approach to
analyze the effects of the hard cutoff on the topological
characteristics. We grow the network by introducing new nodes one by
one for simplicity. Each new node links to $m$ earlier nodes in the
network. The probability that the new node attaches to a previous
node of degree $k$ is defined to be $A_k/A$, where $A_k$ is the rate
of attachment to a previous node and this rate depends only on the
degree of the target node, while $A = \sum_{k=m}^{k_c-1} A_kN_k$ is
the total rate for all events, and $N_k$ is the number of nodes of
degree $k$ in the network. Thus $A_k/A$ equals to the probability
for the newly-introduced node to attach to a node of degree $k$. The
new feature that we study is the effect of a hard cutoff on the
degree of each node. Once the degree of a node reaches $k_c$, it is
defined to become inert so that no further attachment to this node
can occur. Thus only nodes with degrees $k = m, m+1,...,k_c-1$ are
active. This restriction is the source of the cutoff in the
definition of the total attachment rate. We now study the degree
distribution, $N_k(N)$, as a function of the cutoff $k_c$ and the
total number of nodes in the network $N$.

The master equations for the degree distribution
can be written by using the fact that $N_k$ is proportional to $N$,
and thus $N_k \to Nn_k$ as well as $A \to \nu N$ as

\begin{equation}
n_k = \left\{
\begin{array}{ll}
 -\frac{mn_m}{\nu}+1 & \mbox{$k=m$} \\
 \frac{(k-1)n_{k-1}-kn_k}{\nu} & \mbox{$k=m+1,...,k_c-1$} \\
  \frac{(k_c-1)n_{k_c-1}}{\nu} & \mbox{$k=k_c$}
    \end{array} \right.
\;.
\end{equation}
By the nature of these equations, it is evident that $n_{k_c}$ is of
a different order than $n_k$ with $k < k_c$. Starting with the
solution $n_m = \nu/(m+\nu)$, we can find $n_k$ by subsequent
substitutions. This recursive approach gives us a chance to write
$n_k$ values as products \cite{KRAPIVSKY01} and by converting these
products into Euler gamma functions we show that $n_{k_c}$ scales as
$k^{-\nu}$, while for $k < k_c$, $n_k$ scales as $k^{-(\nu+1)}$. We
can obtain the coefficient $\nu$ in $A = \nu N$ self consistently
from $A = \sum_{k=m}^{k_c-1} A_kn_k \equiv \nu N$, or equivalently,
$\nu = \sum_{k=m}^{k_c-1} A_kn_k$.  By rewriting the sum above as a
difference between two sums with limits from the minimum degree to
$\infty$ and from cutoff to $\infty$ and by taking asymptotic limits
\cite{CONCRETE_MATH} of large $N$ and $k_c$ we get

\begin{equation}
\nu \to 2-\frac{2m}{k_c}
\;.
\label{nu_4}
\end{equation}
This result shows that $n_k \sim k^{-(3-2m/k_c)}$ for $k<k_c$ and
$n_{k_c} \sim k_c^{-(2-2m/k_c)}$ confirming the change in the degree
distribution exponent \cite{guclu-2007}. This implies that any finite
hard cutoff value decreases the degree distribution exponent, i.e.,
it makes the degree distribution flatter. A better search efficiency
observed for a smaller cutoff can be explained by the increase in the
degree distribution exponent \cite{guclu-2007}.

Ad-hoc scale-free networks have recently attracted considerable
attention in the literature mainly because of its most-desired
property of robustness to random attacks or failures. For example it
was shown that \cite{alb00} the diameter of the Internet at the
autonomous system level, which is the most famous example of
scale-free networks, would not be changed considerably if up to
2.5\% of the routers were removed randomly. This is an order of
magnitude larger than the failure rate. It was also shown in
\cite{alb00} that for a scale-free network of size 10,000 and a
failure rate of 18\%, the biggest connected component holds 8,000
nodes, whereas under the same conditions a random network can
survive this failures by the biggest connected component of size
100.

Many models for ad-hoc scale-free networks in which the edges can
appear and disappear \cite{doro00,krap02,doro01} or some nodes are
removed \cite{roy04} have been studied. In the first set of studies
as the nodes are joining to the network some links among the
pre-existing nodes are rewired or moved randomly by some probability
parameter. Depending on the parameters such models exhibit either
exponential or power-law degree distributions. In \cite{roy04}, as
the nodes are joining by preferential attachment some randomly
selected nodes are deleted along with their links from the network.
If the nodes whose neighbors are deleted do not reconnect themselves
to other nodes, it is observed that the degree distribution is a
power law with an exponent ranging from 3 to infinity depending on
the deletion probability. The authors proposed a remedy for the
deletion that the neighbors of the deleted nodes select some other
nodes in the network and connect by again using preferential
attachment rules. In this case the degree distribution is still a
power law but the exponent changes from 3 to 2 as the deletion
probability goes from 0 to 1.

The main disadvantage of the ad-hoc scale-free models in the
literature is that they lack localized algorithmic solutions. All
requires global information to be available to nodes so that they
can reconnect to randomly selected nodes in the network. For this
reason we grow scale-free networks with local heuristics only to
simulate the real-life situation in unstructured peer-to-peer
systems. To parameterize our model we use two different time-to-live
variables: $\tau_j$ and $\tau_l$ to describe the number of nodes
available to a new node and to a neighbor of a deleted node,
respectively. In the next section we explain our model and its
parameters in detail.

\begin{figure}
\centering
\includegraphics[keepaspectratio=true,angle=0,width=80mm]{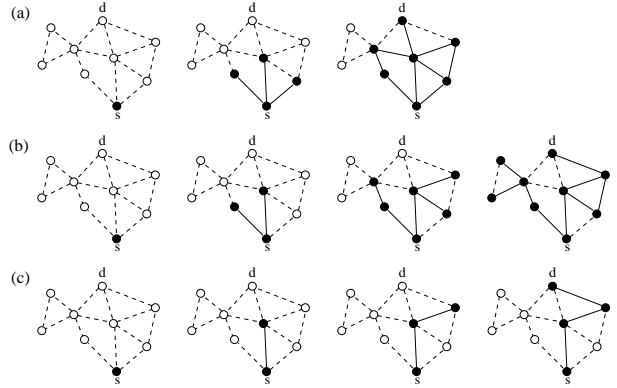}
\vspace{-3mm} \caption{Search strategies: (a) Flooding (b)
Normalized flooding (c) Random walk} \vspace{-7mm}
\label{fig_search}
\end{figure}

\begin{figure*}
\begin{center}
\begin{tabular}{ccc}
\hspace{-4mm}
\includegraphics[keepaspectratio=true,angle=0,width=59mm]
{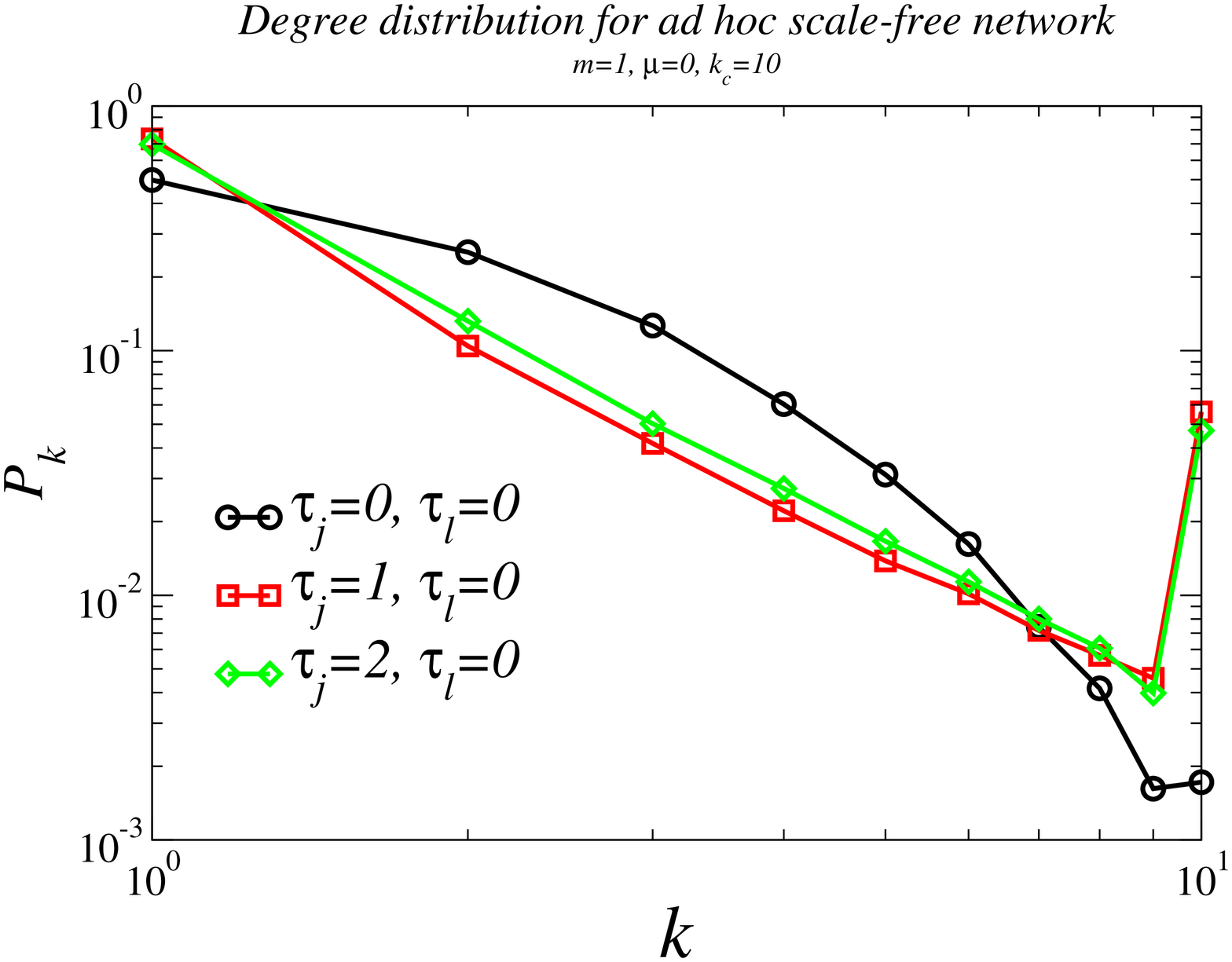} & \hspace{-5mm}
\includegraphics[keepaspectratio=true,angle=0,width=59mm]
{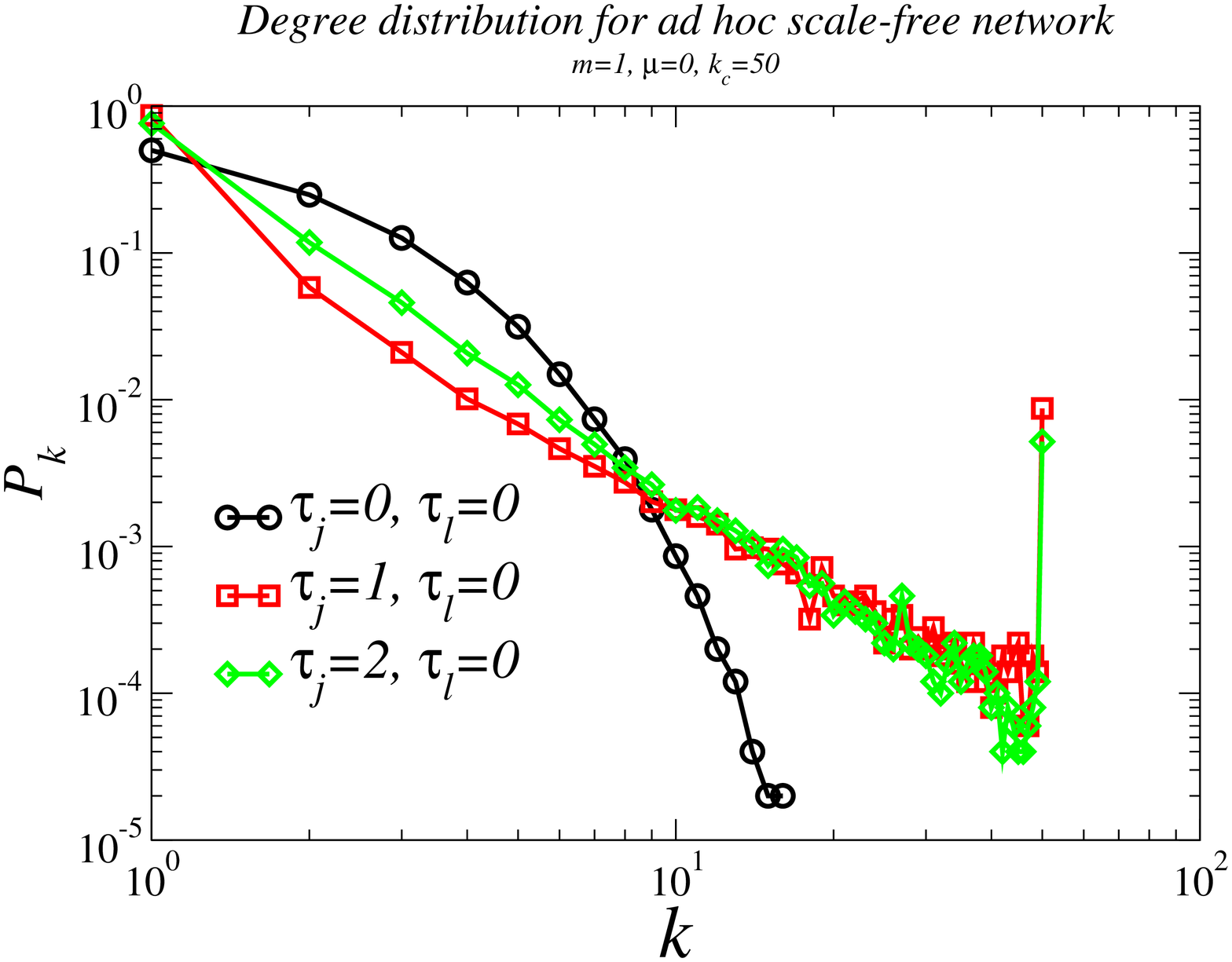} & \hspace{-5mm}
\includegraphics[keepaspectratio=true,angle=0,width=59mm]
{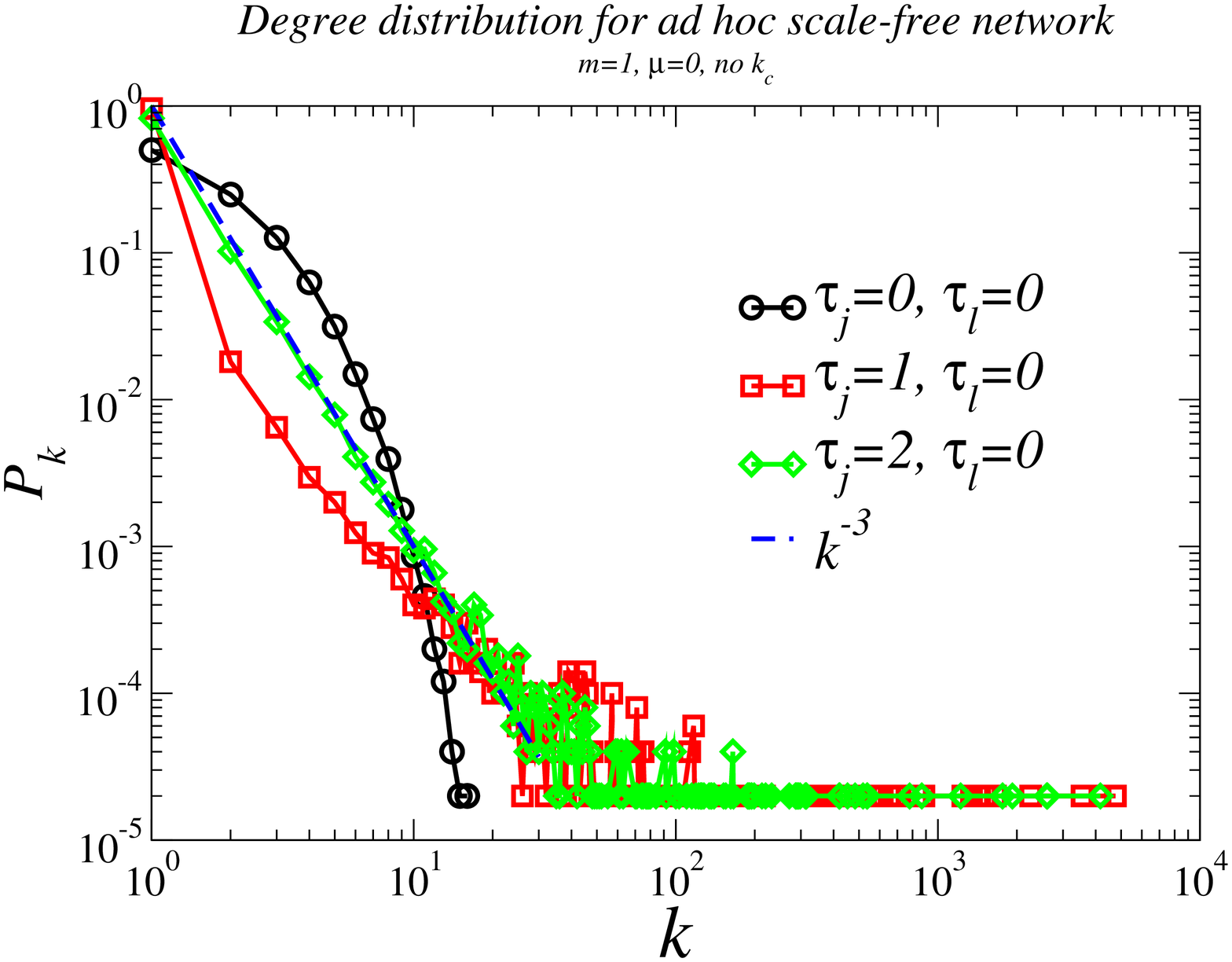} \vspace{-2mm} \\
\small{(a) $m$=1, $k_c$=10} & \small{(b) $m$=1, $k_c$=50} &
\small{(c) $m$=1, no cutoff} \vspace{2mm} \\
\hspace{-4mm}
\includegraphics[keepaspectratio=true,angle=0,width=59mm]
{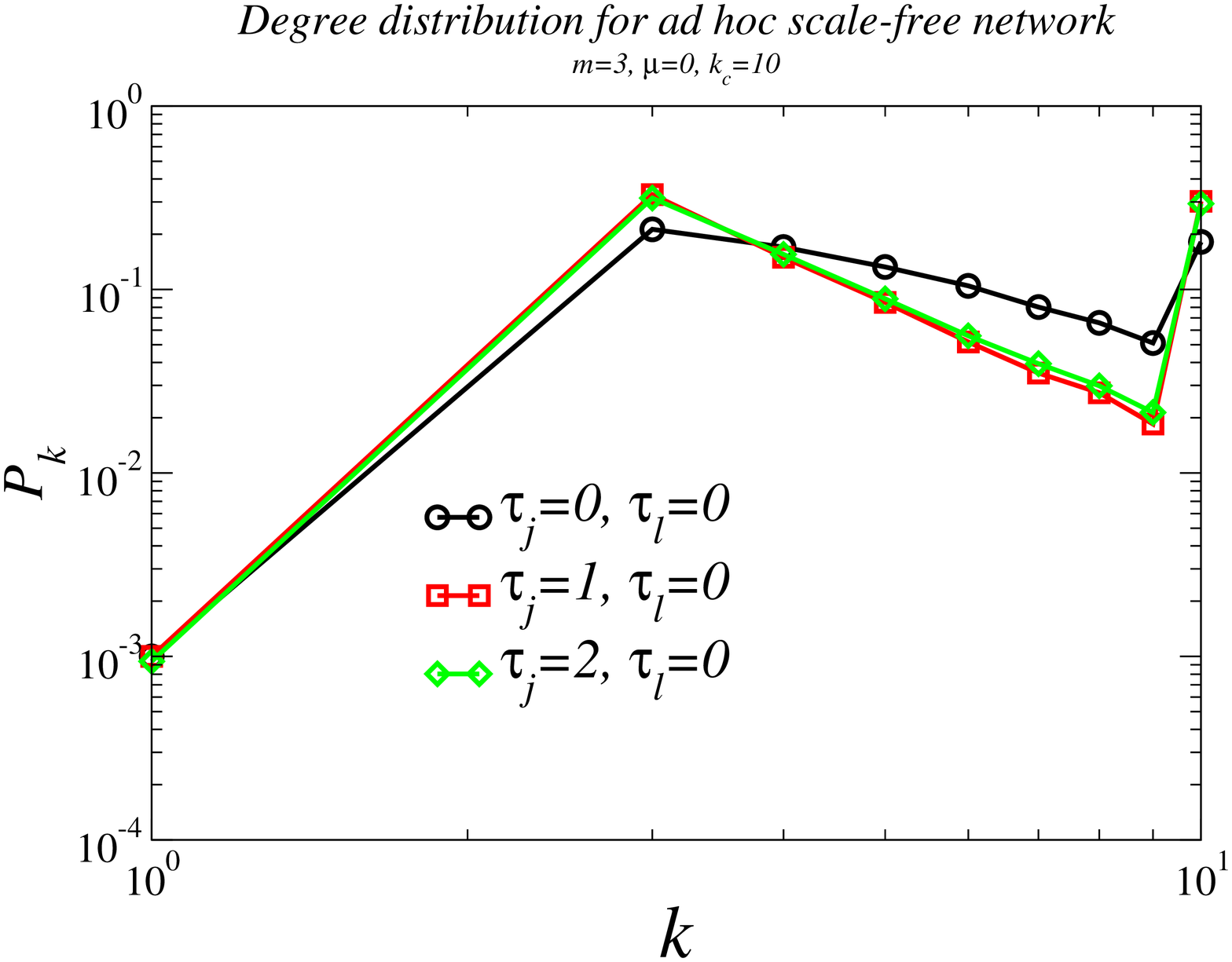} & \hspace{-5mm}
\includegraphics[keepaspectratio=true,angle=0,width=59mm]
{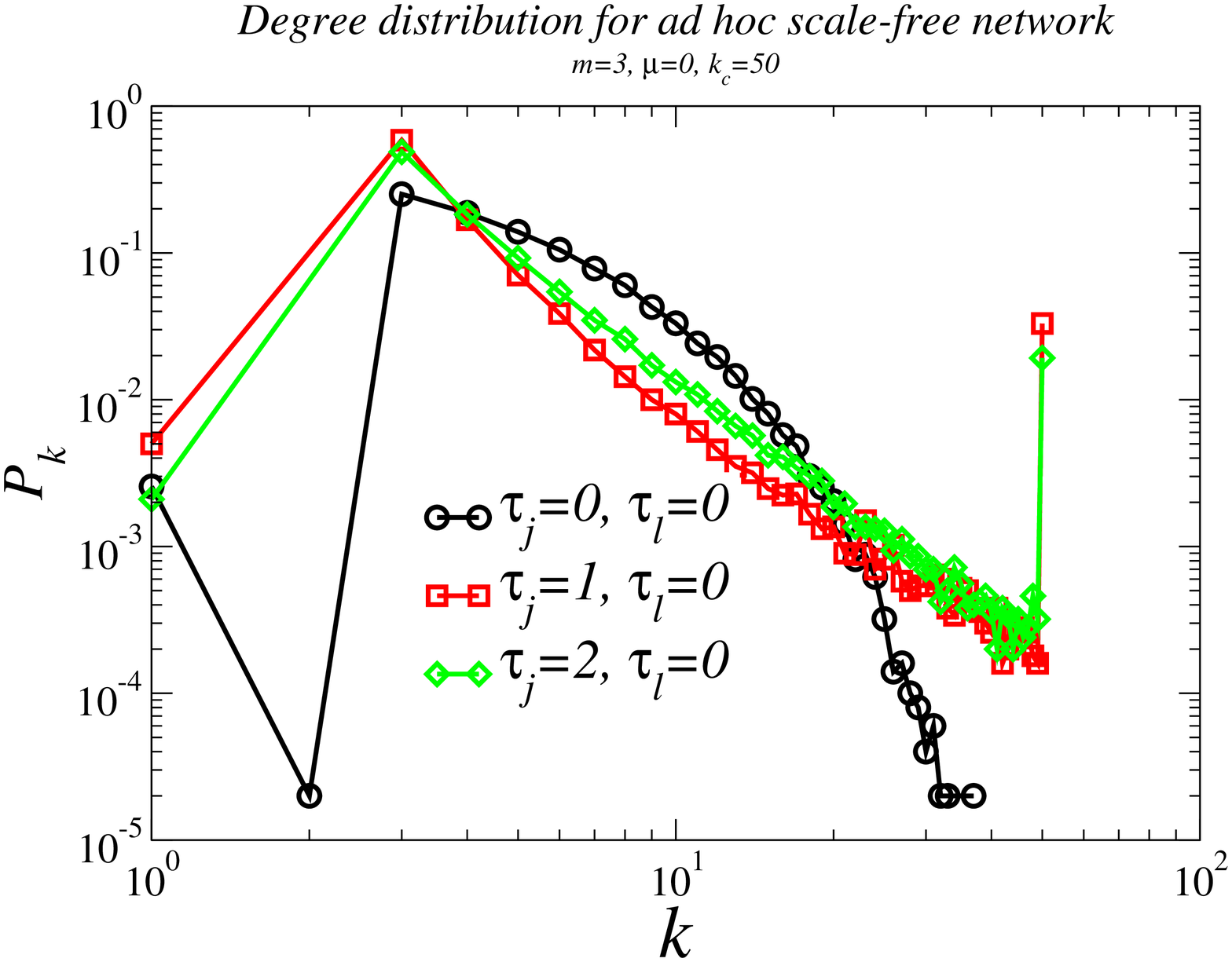} & \hspace{-5mm}
\includegraphics[keepaspectratio=true,angle=0,width=59mm]
{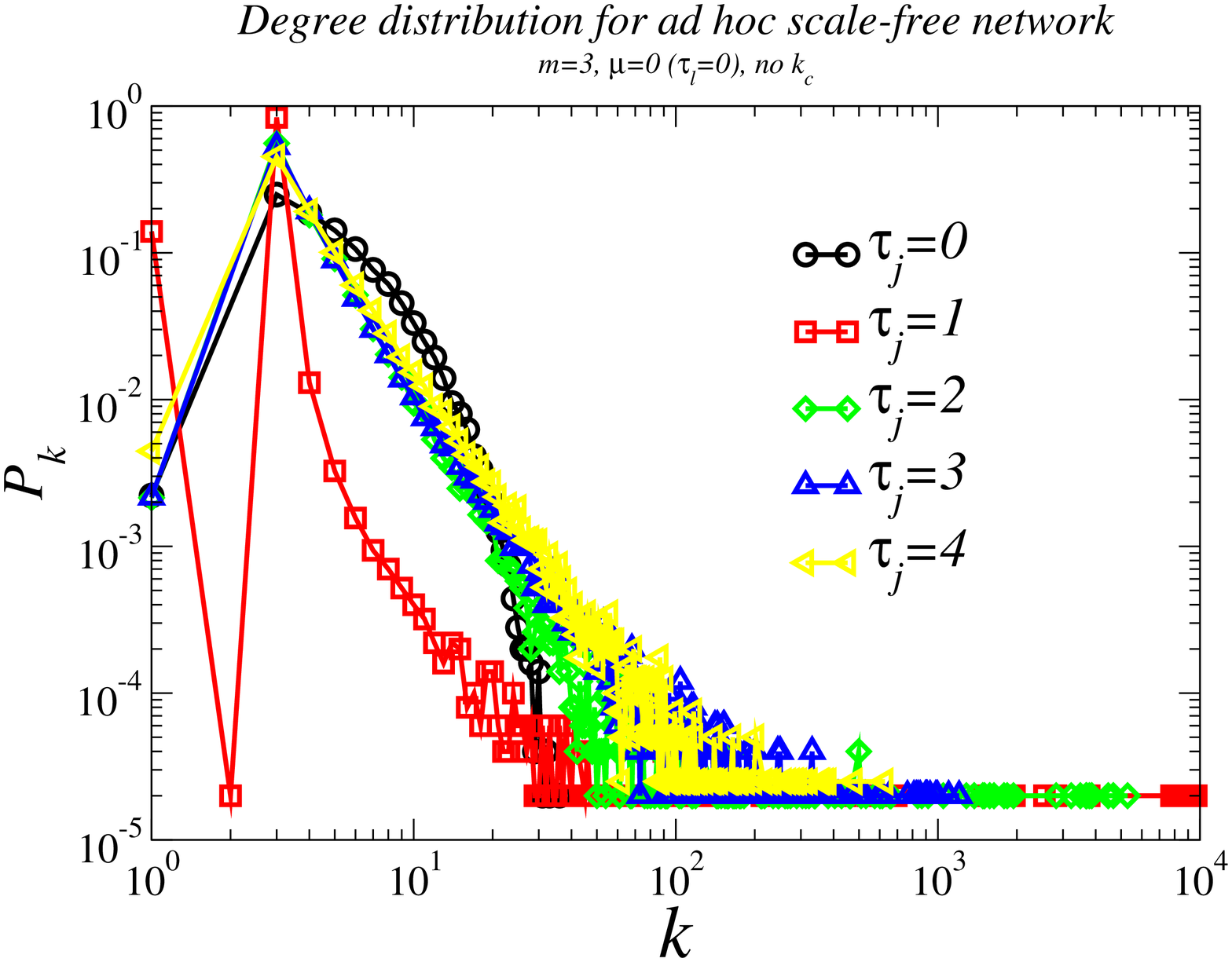} \vspace{-2mm} \\
\small{(d) $m$=3, $k_c$=10} & \small{(e) $m$=3, $k_c$=50} &
\small{(f) $m$=3, no cutoff}
\end{tabular}
\end{center}
\vspace{-6mm} \caption{Degree distributions when there is no
ad-hocness (i.e., $\mu$=0): $P(k)$ for various networks generated by
our framework for varying $\tau_j$.} \label{fig:pk-mu=0}
\vspace{-8mm}
\end{figure*}

\Section{Growing Scale-Free Topologies with Local Heuristics}
\label{sec:local-heuristics}

In the PA model and its ad-hoc variants as outlined in the previous
section, the new node or the neighbor of a deleted node has to make
random attempts to connect to the existing nodes with a probability
depending on the degree of the existing nodes. To implement this in
a P2P (or any distributed) environment, the new node has to have
information about the global topology (e.g. the current number of
degrees each node has for the PA model), which might be very hard to
maintain in reality. Thus, in order for a topology construction
mechanism to be practical in P2P networks, it must allow joining or
rewiring of the nodes by just using locally available information.
Of course, the cost of using only local information is expected to
be loss of scale-freeness (or any other desired characteristics) of
the whole overlay topology, which will result in loss of search
efficiency in return. In this section, we present a practical method
using local heuristics and no global information about the topology.
This model imitates the method for finding peers in Gnutella-like
unstructured P2P networks.

In our model, starting with some $m+1$ fully connected nodes, at
every time step a new node with $m$ possible links is added to the
network and one randomly chosen node is deleted with probability
$\mu$. Since the nodes are short-sighted, i.e., they do not have
global information about the network, they can only choose from a
subset of the network (\textit{horizon}) they construct instead of
the whole network. The parameter $\tau_j$ and $\tau_l$ are the TTL
values used by the nodes and denote the measure of locality in
joining and leaving, respectively. A newly added node, first, select
a random existing node and construct a set of nodes reachable in
$\tau_j$ hops or less from that node. Then, this new node randomly
selects a node from this set and connects itself with probability
proportional to its degree. This probability is normalized by the
total degree of the nodes in the set. The new node randomly selects
other nodes in the set until its degree reaches $m$. If no node is
left in the set to connect but the degree of the new node is less
than $m$, it selects another random node from the network and
continue this process. In the deletion case, the neighbors of the
deleted node selects a node randomly from a set of nodes reachable
in $\tau_l$ or less steps from the deleted node and connect by using
the preferential attachment rule. Here, in both cases nodes cannot
connect to other nodes with degrees equal to the hard cutoff.

There are special cases in this model: i) when $\tau_j=0$, the
horizon of the new nodes contain only the randomly selected nodes
and the preferential attachment rule is invalid. In this case the
new node connect to this single node in the horizon if its degree is
less than the hard cutoff. ii) when $\tau_l=0$ the neighbors of the
deleted node do not have any node in their horizons so no rewiring
occurs. These nodes just loses one of their links and they do
nothing to compensate it. The model typically becomes the
preferential attachment with global information when $\tau_j$ value
is large and $\tau_l$ is zero and a BA network with $\gamma=3$ is
obtained.

\Section{Simulations} \label{sec:simulations}

In the previous sections, we introduced a framework to investigate
the effects of join and leave processes in terms of scale-freeness
of the topology being constructed within the context of ad-hoc
unstructured P2P networks. Here, we study a number of
message-passing algorithms that can be efficiently used to search
items in P2P networks utilizing the scale-free degree distribution
in sample networks generated by our topology construction
algorithms. These search algorithms are completely decentralized and
do not use any kind of global knowledge about the network. We
consider three different search algorithms: \emph{flooding} (FL),
\emph{normalized flooding} (NF), and \emph{random walk} (RW).

Goals of our simulation experiments include:
\begin{itemize}
\item{\emph{Effect of ad-hocness on the search efficiency in an
uncooperative environment with hard cutoffs:}} Ad-hocness of nodes
joining or leaving the network affects the search efficiency, i.e.,
\emph{number of hits per unit time}. Further, applying hard cutoffs
on such ad-hoc scale-free topologies reduces the degree distribution
exponent. We are interested in observing the effect of this
ad-hocness and hard cutoffs on the search efficiency for three
search algorithms, i.e., FL, NF, and RW. This extends our previous
work in \cite{guclu-2007}, which focussed on the effect of hard
cutoffs \emph{only}.

\item{\emph{Ad-hoc scale-free topology construction with global vs. local
information:}} Though we showed in the previous section that using
local information when a peer is joining yields a less scale-free
topology, the effect of this on search efficiency still needs to be
shed light on. Our simulations aim to investigate this too.
\end{itemize}

\begin{figure*}
\begin{center}
\begin{tabular}{ccc}
\hspace{-4mm}
\includegraphics[keepaspectratio=true,angle=0,width=58mm]
{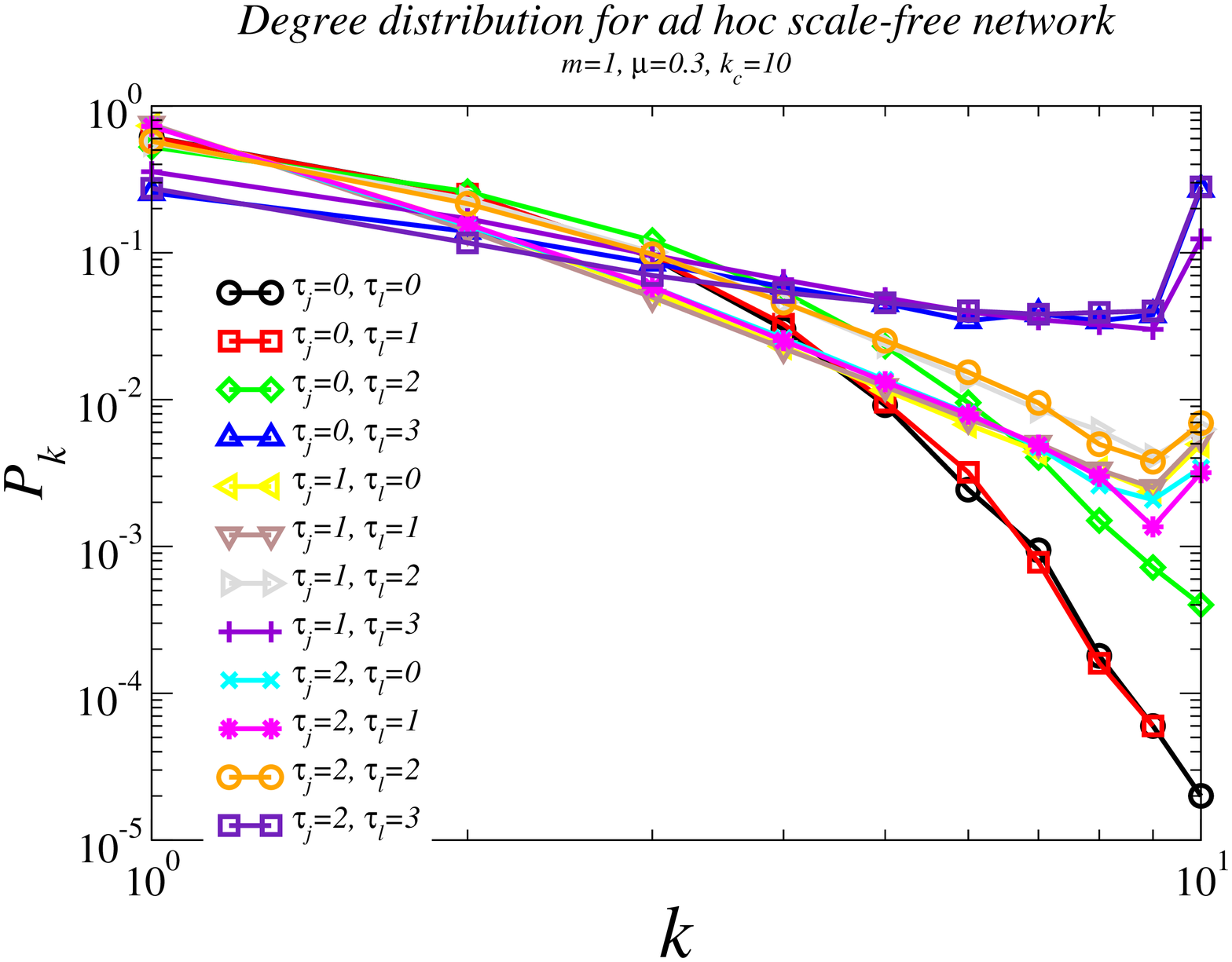} & \hspace{-5mm}
\includegraphics[keepaspectratio=true,angle=0,width=58mm]
{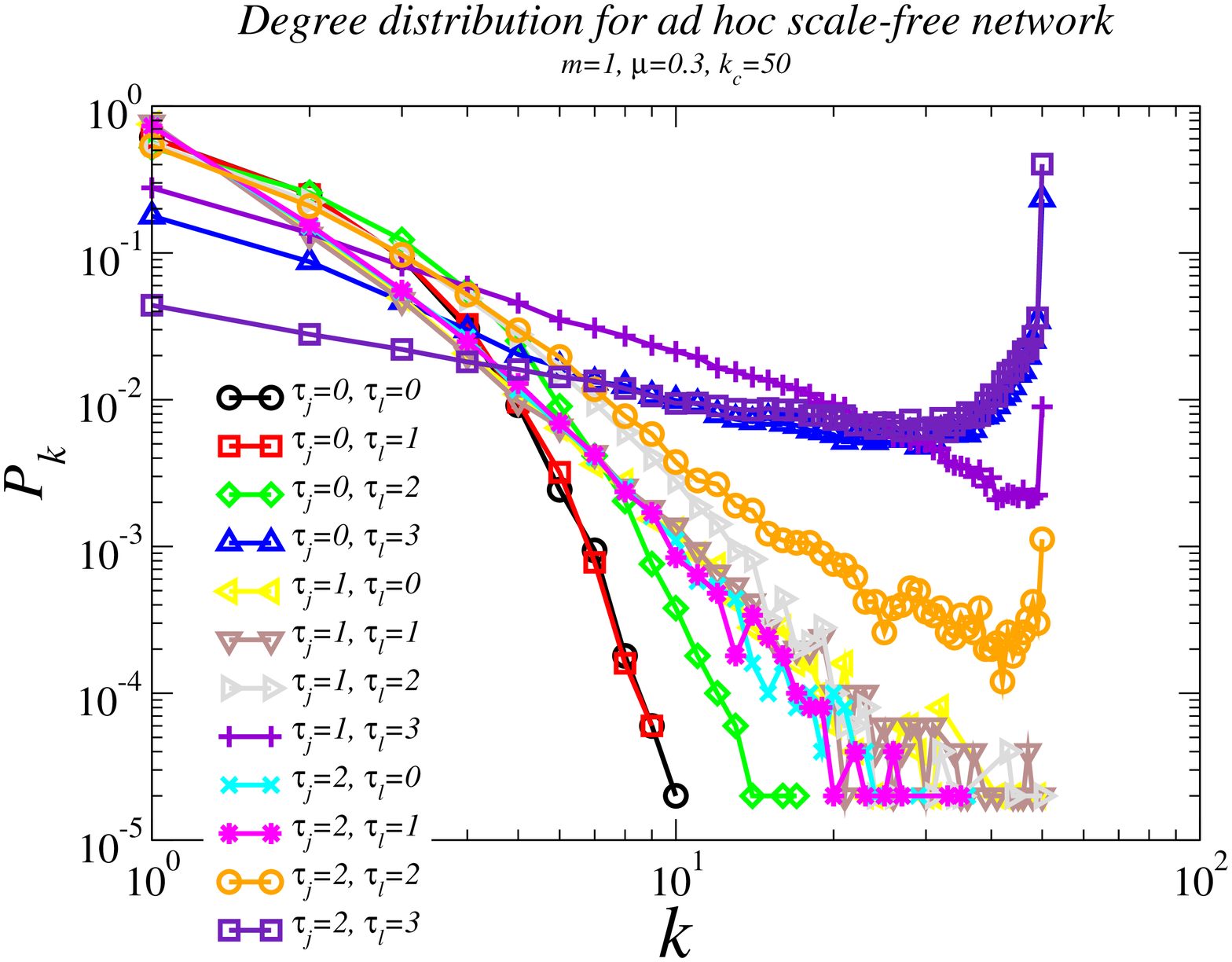} & \hspace{-5mm}
\includegraphics[keepaspectratio=true,angle=0,width=59mm]
{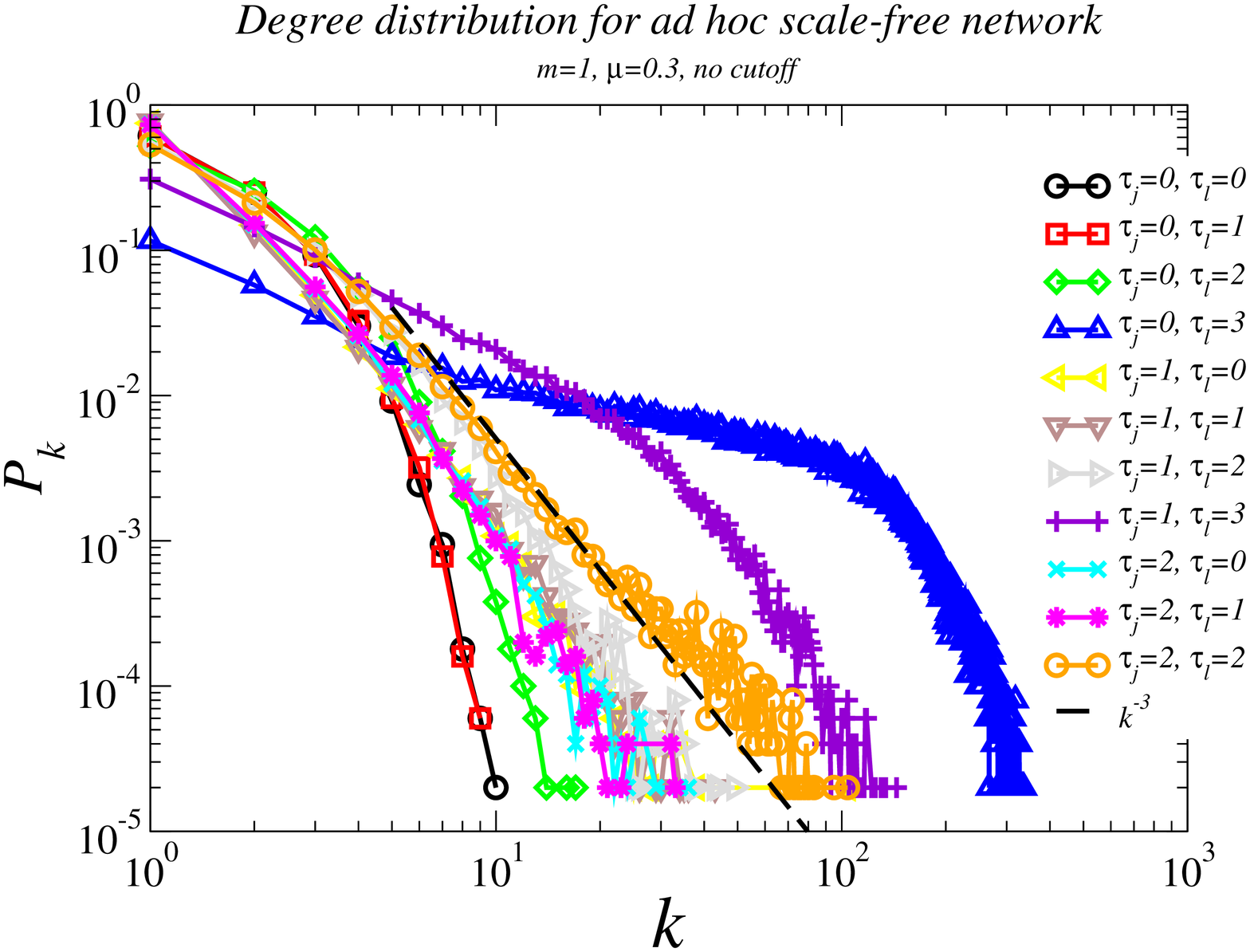} \vspace{-2mm} \\
\small{(a) $m$=1, $k_c$=10} & \small{(b) $m$=1, $k_c$=50} &
\small{(c) $m$=1, no cutoff} \\
\end{tabular}
\end{center}
\vspace{-6mm} \caption{Degree distributions over ad-hoc nodes (i.e.,
$\mu$=0.3): $P(k)$ for various networks generated by our framework
for varying $\tau_j$ and $\tau_l$.} \label{fig:pk-mu=0.3}
\vspace{-8mm}
\end{figure*}

\SubSection{Search Algorithms}

We use three search techniques to evaluate our ad-hoc scale-free
topologies:

\vspace{2mm} \noindent \textbf{Flooding (FL):} FL is the most common
search algorithm in unstructured P2P networks. In search by FL, the
source node $s$, sends a message to all its nearest neighbors. If
the neighbors do not have the requested item, they send on to their
nearest neighbors excluding the source node [see
Fig.~\ref{fig_search}(a)]. This process is repeated a certain number
of times, which is usually called \textit{time-to-live (TTL)}.


\vspace{2mm} \noindent \textbf{Normalized Flooding (NF):}
In NF, the minimum degree $m$ in the network is an important factor.
NF search algorithm proceeds as follows: When a node of degree $m$
receives a message, the node forwards the message to all of its
neighbors excluding the node forwarded the message in the previous
step. When a node with larger degree receives the message, it
forwards the message only to randomly chosen $m$ of its neighbors
except the one which forwarded the message. The NF mechanism is
illustrated in Fig.~\ref{fig_search}(b). In this simple network with
$m=2$, the source node sends a message to its randomly chosen two
neighbors and these neighbors forward the message to their randomly
chosen two neighbors. In the third step, the message reaches its
destination.

\vspace{2mm} \noindent \textbf{Random Walk (RW):} RW or multiple RWs
have been used as an alternative search algorithm to achieve even
better granularity than NF. In RW, the message from the source node
is sent to a randomly chosen neighbor. Then, this random neighbor
takes the message and sends it to randomly selected one of its
random neighbors excluding the node from which it got the message.
This continues until the destination node is reached or the total
number of hops is equal to TTL. A schematic of RW can be seen in
Fig.~\ref{fig_search}(c). RW can also be seen as a special case of
FL where only one neighbor is forwarded the search query, providing
the other extreme situation of the tradeoff between delivery time
and messaging complexity.

\renewcommand{\baselinestretch}{1}
\begin{table}
    \caption{Parameters of our topology construction framework}
    \label{tab:parameters}
    \begin{center}
\vspace{-5mm}
    \begin{tabular}{|c|c|c|}
    \hline
        Symbol & Parameter Description & Range \\
    \hline
        $\mu$ & Ad-hocness of the nodes & [0,1) \\
        $\tau_j$ & Available information at join & $\ge 0$ \\
        $\tau_l$ & Available information at leave & $\ge 0$ \\
        $k_c$ & Hard cutoff & $\ge 1$ \\
        $m$ & Minimum degree (\# of stubs) & $\ge 1$ \\
    \hline
    \end{tabular}
\vspace{-7mm}
    \end{center}
\end{table}

\begin{figure}
\begin{center}
\includegraphics[keepaspectratio=true,angle=0,width=84mm]
{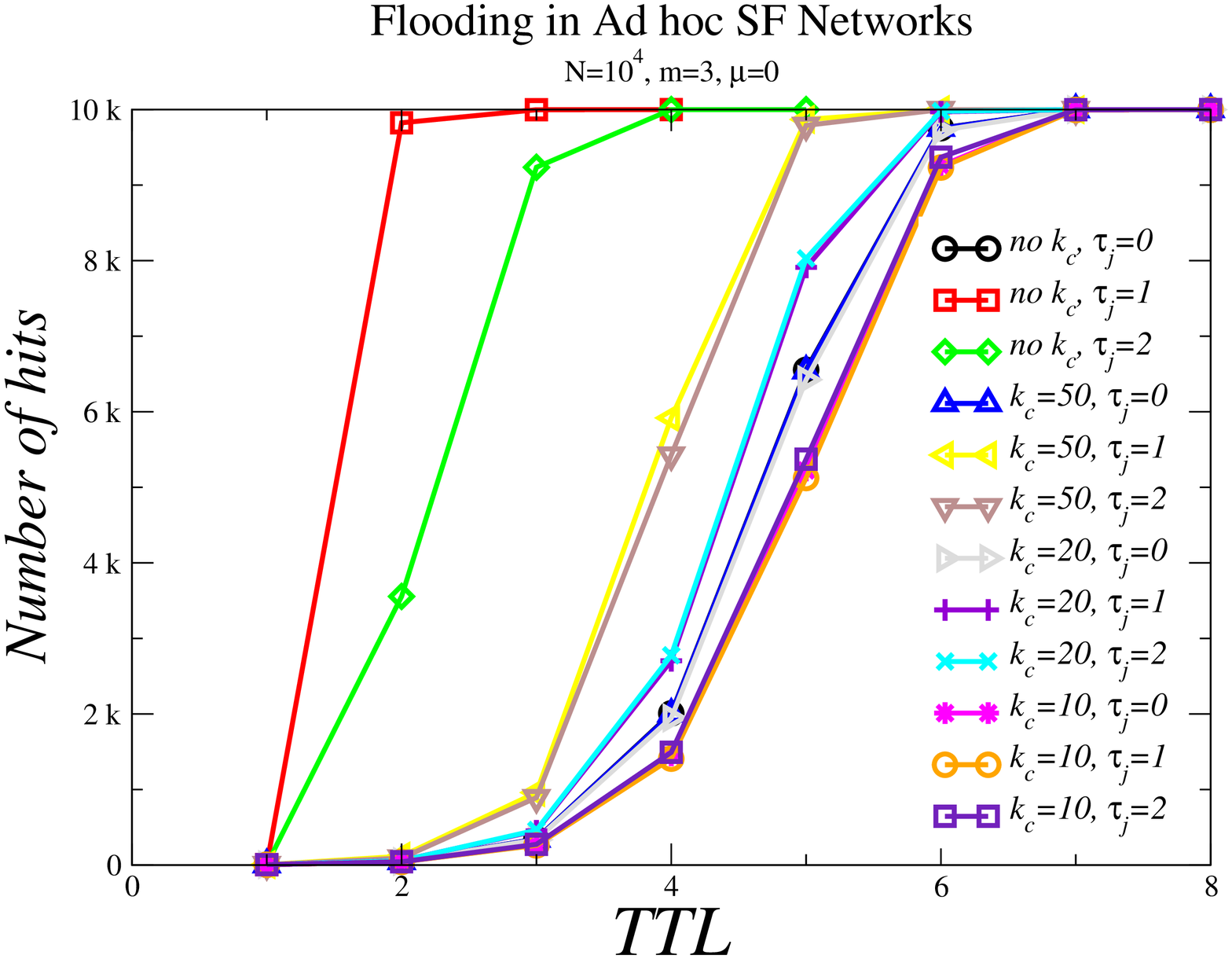}
\end{center}
\vspace{-7mm} \caption{Flooding (FL) performance over topologies
with $m$=3 and no ad-hocness.} \label{fig:FL1} \vspace{-11mm}
\end{figure}

\SubSection{Results}

We simulated the three search algorithms FL, NF, and RW on the
topologies generated by our framework with three different
parameters: (i) ad-hocness, $0< \mu < 1$, (ii) available information
during join, $\tau_j \ge 0$, (iii) available information during
leave, $\tau_l \ge 0$, (iv) hard cutoff, $k_c > 1$, and (v) minimum
degree (number of stubs), $m$. These parameters are listed in
Table~\ref{tab:parameters} as well. By assigning different values to
each of these parameters, we generated topologies with 10000 nodes.
We used different $k_c$ values from 10 to 100 (or just a few in this
range), in addition to the natural cutoff, i.e., no hard cutoff. We
varied $\tau_j$ and $\tau_l$ from 0 to 3. Minimum degree values (or
$m$) in our topologies were 1, 2, or 3. We studied smaller values of
$\mu$ from 0 to 0.3, reflecting no churn to 30\% churn,
respectively. We performed 5 realizations of our results.

We varied the TTL (i.e., time-to-live) values of search queries in
FL and NF to the point we reach the system size. To compare search
efficiencies of RW and NF fairly, we equated TTL of RW searches to
the number of messages incurred by the NF searches in the same
scenario. Thus, for the search efficiency graphs of RW when TTL is
equal to a particular value such as 4, this means that the number of
hits corresponding to that TTL=4 value is obtained by simulating a
RW search with TTL equal to the number of messages that were caused
by an NF search using a TTL value of 4. A similar normalization was
done in \cite{GMS05}.

\subsubsection{Effects on Degree Distribution}
Our simulation results show the effect of ad-hocness and hard cutoff
on the degree distribution of the topologies.
Figure~\ref{fig:pk-mu=0} shows the degree distribution of the
topologies when there is no ad-hocness, i.e., $\mu$=0. Similarly,
Figure~\ref{fig:pk-mu=0.3} shows the degree distributions when the
nodes are ad-hoc with $\mu$=0.3. It is known that using more global
information (i.e., knowing more of the network topology) helps to
generate better scale-free topologies with lower power-law exponent.
This phenomenon is clearly shown in Figure~\ref{fig:pk-mu=0}, i.e.,
the degree distribution shifts from an Exponential one to a
power-law one as $\tau_j$ increases from 0 to 2 (i.e., the joining
node uses the topology information at a larger horizon). This is
true for both $m$=3 and $m$=1, though larger $m$ makes the shift a
little less apparent. Further, the hard cutoff, $k_c$, only affects
this process by simply bounding the very large hubs to the cutoff
without affecting the transition from Exponential to power-law.

An interesting result being revealed in Figure~\ref{fig:pk-mu=0.3}
is that $\tau_l$ has much more significant effect than $\tau_j$ in
shifting the degree distribution from Exponential to power-law. This
is even more apparent for smaller values of the cutoff.

\subsubsection{Effects on Search Efficiency}

In flooding by far the most important parameter when there is no
deletion in the network is the cutoff which determines the number of
distinct nodes one can reach from a node, see Figure~\ref{fig:FL1}.
In this case, $\tau_j$ is also an important parameter which changes
the network from an exponential to a scale-free one and give better
efficiency in flooding.

Our simulations also show that this effect can be relieved by
increasing the minimum degree in the network as it can be seen in
Figure~\ref{fig:FL2}. More interestingly, ad-hocness plays an
important role in the efficiency of search algorithms. Negative
effect of the high ad-hocness (high $\mu$) can be eliminated by
increasing the available information in rewiring, i.e., by
increasing $\tau_l$ in both flooding and normalized flooding, see
Figure~\ref{fig:NF}. In some cases in normalized flooding higher
ad-hocness yields better efficiency for enough high values of
$\tau_l$. Here, we do not present results for random walk search
algorithm since the qualitatively they are not different than
normalized flooding except that the random walk is more vulnerable
to isolated clusters in the network.

\begin{figure*}
\begin{center}
\begin{tabular}{ccc}
\hspace{-4mm}
\includegraphics[keepaspectratio=true,angle=0,width=59mm]
{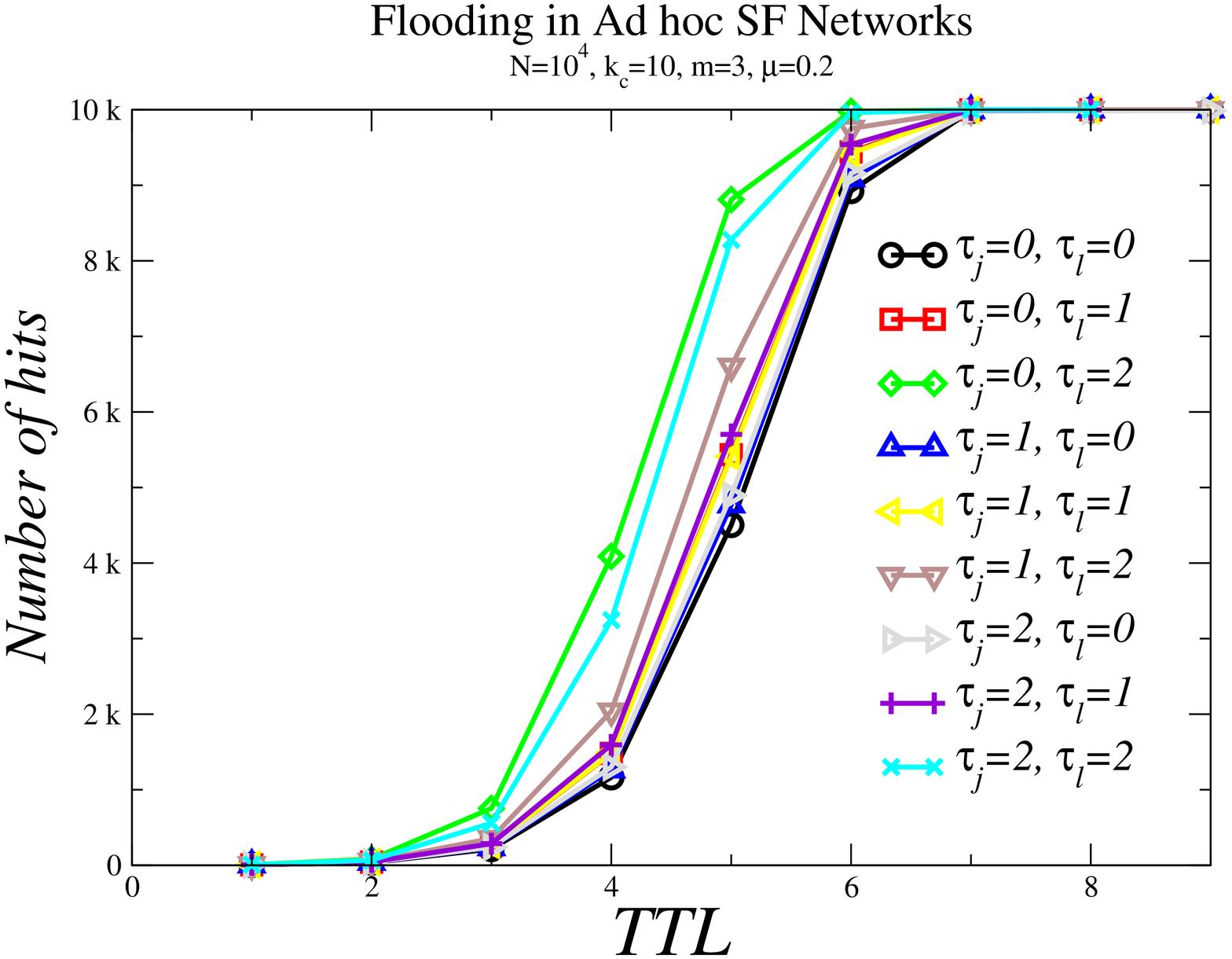} & \hspace{-5mm}
\includegraphics[keepaspectratio=true,angle=0,width=59mm]
{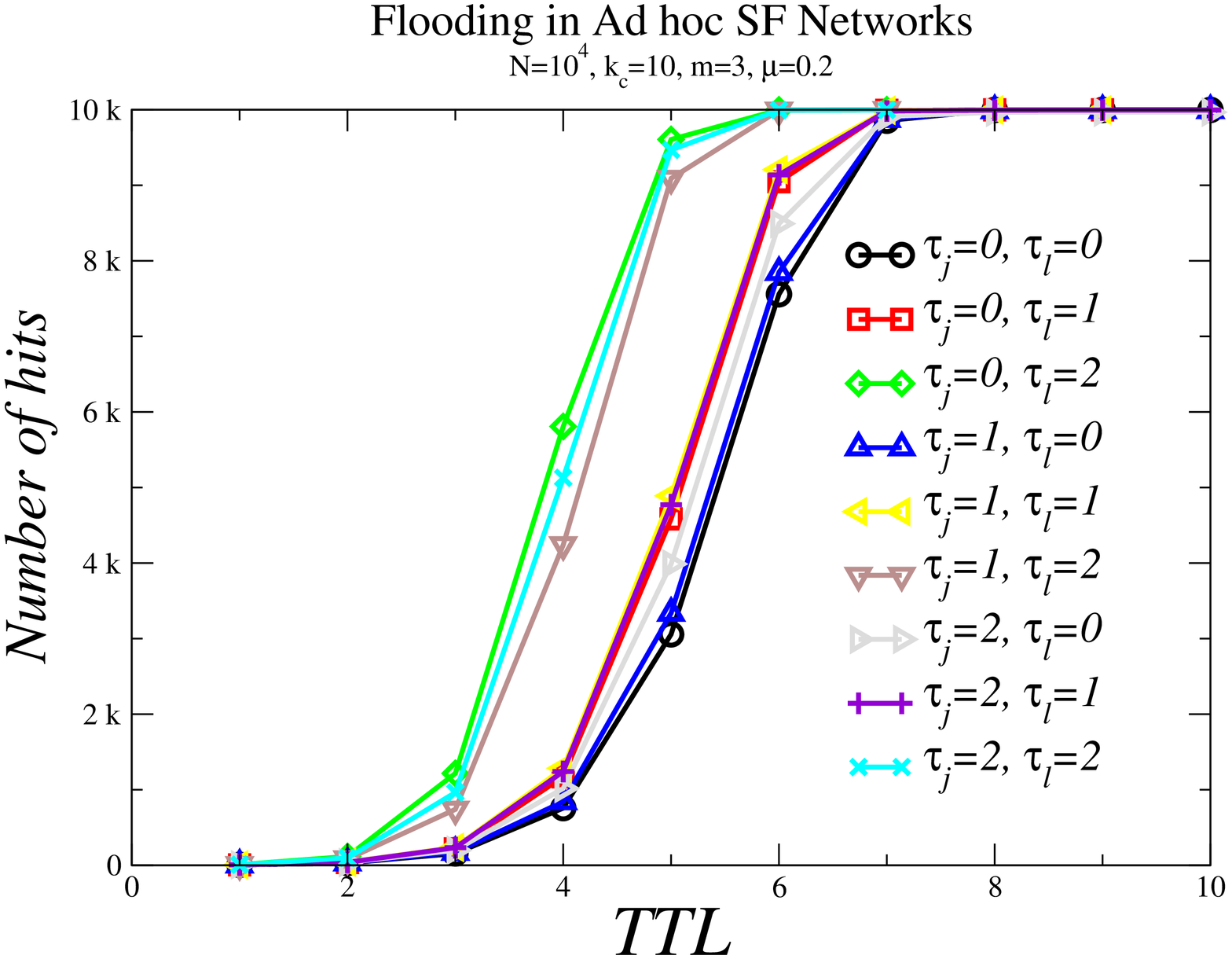} & \hspace{-5mm}
\includegraphics[keepaspectratio=true,angle=0,width=59mm]
{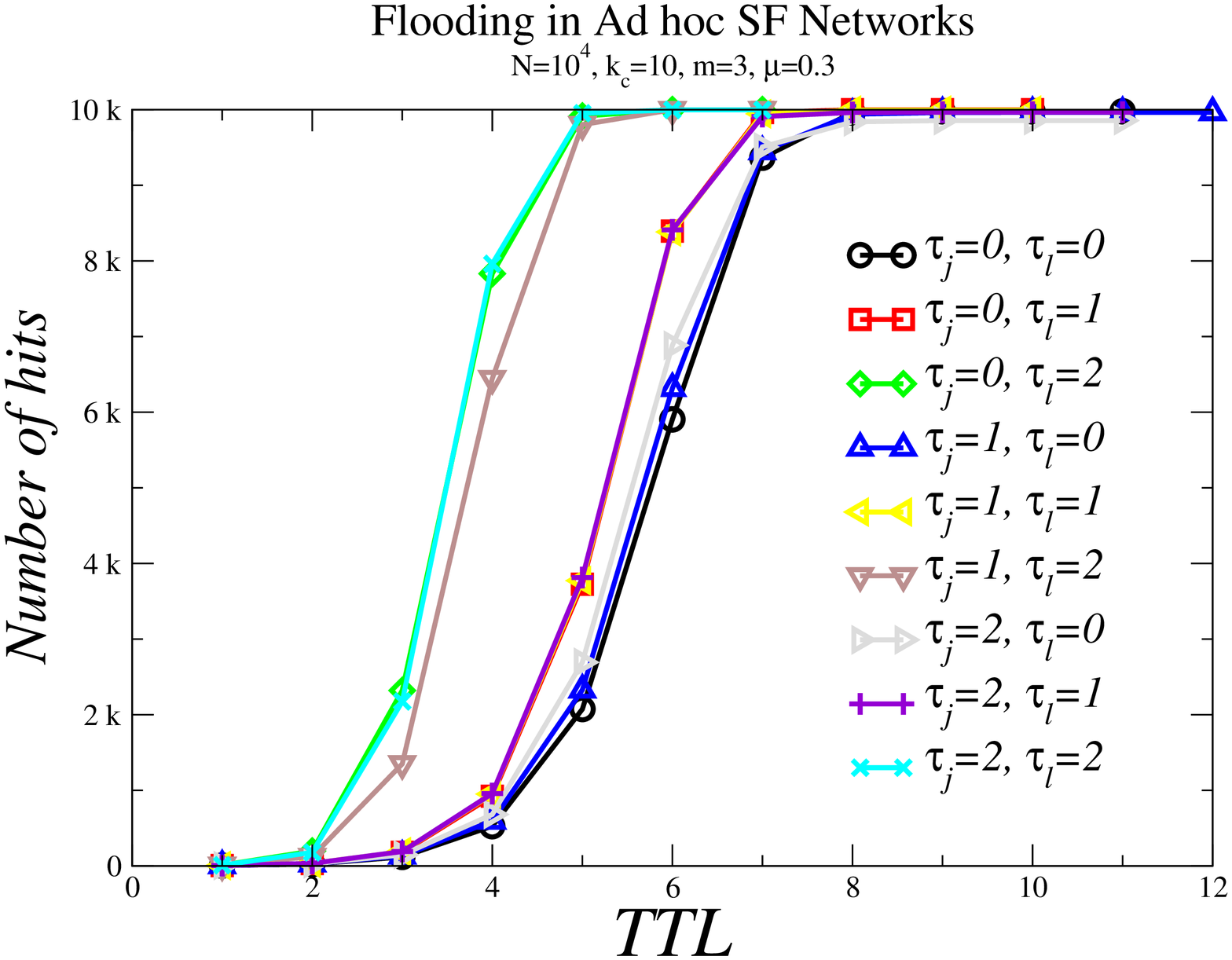} \vspace{-2mm} \\
\small{(a) $\mu$=0.1} & \small{(b) $\mu$=0.2} &
\small{(c) $\mu$=0.3} \\
\end{tabular}
\end{center}
\vspace{-6mm} \caption{Flooding (FL) performance over topologies
generated with $k_c$=10 and $m$=3.} \label{fig:FL2} \vspace{-1mm}
\end{figure*}

\begin{figure*}
\begin{center}
\begin{tabular}{ccc}
\hspace{-4mm}
\includegraphics[keepaspectratio=true,angle=0,width=59mm]
{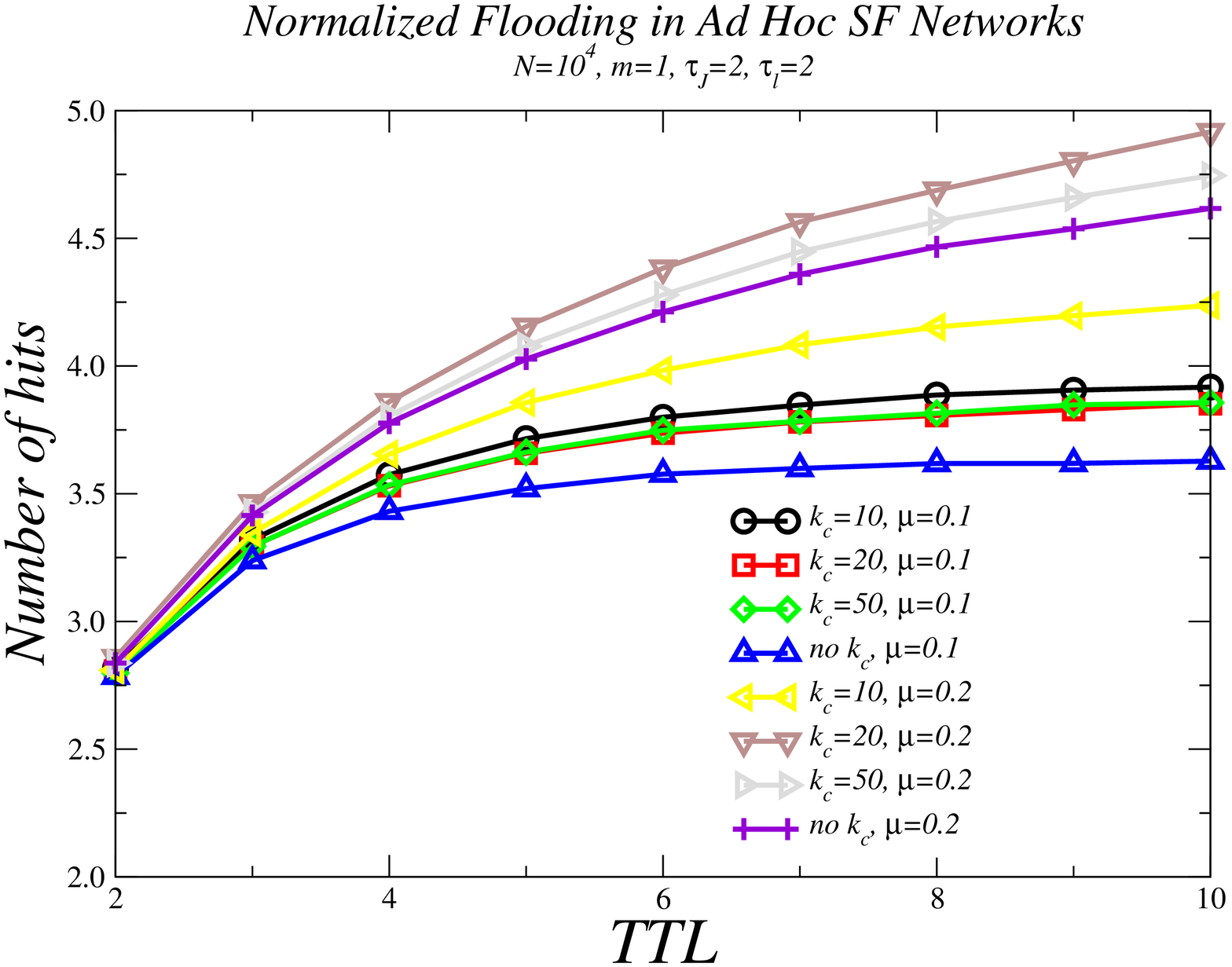} & \hspace{-5mm}
\includegraphics[keepaspectratio=true,angle=0,width=59mm]
{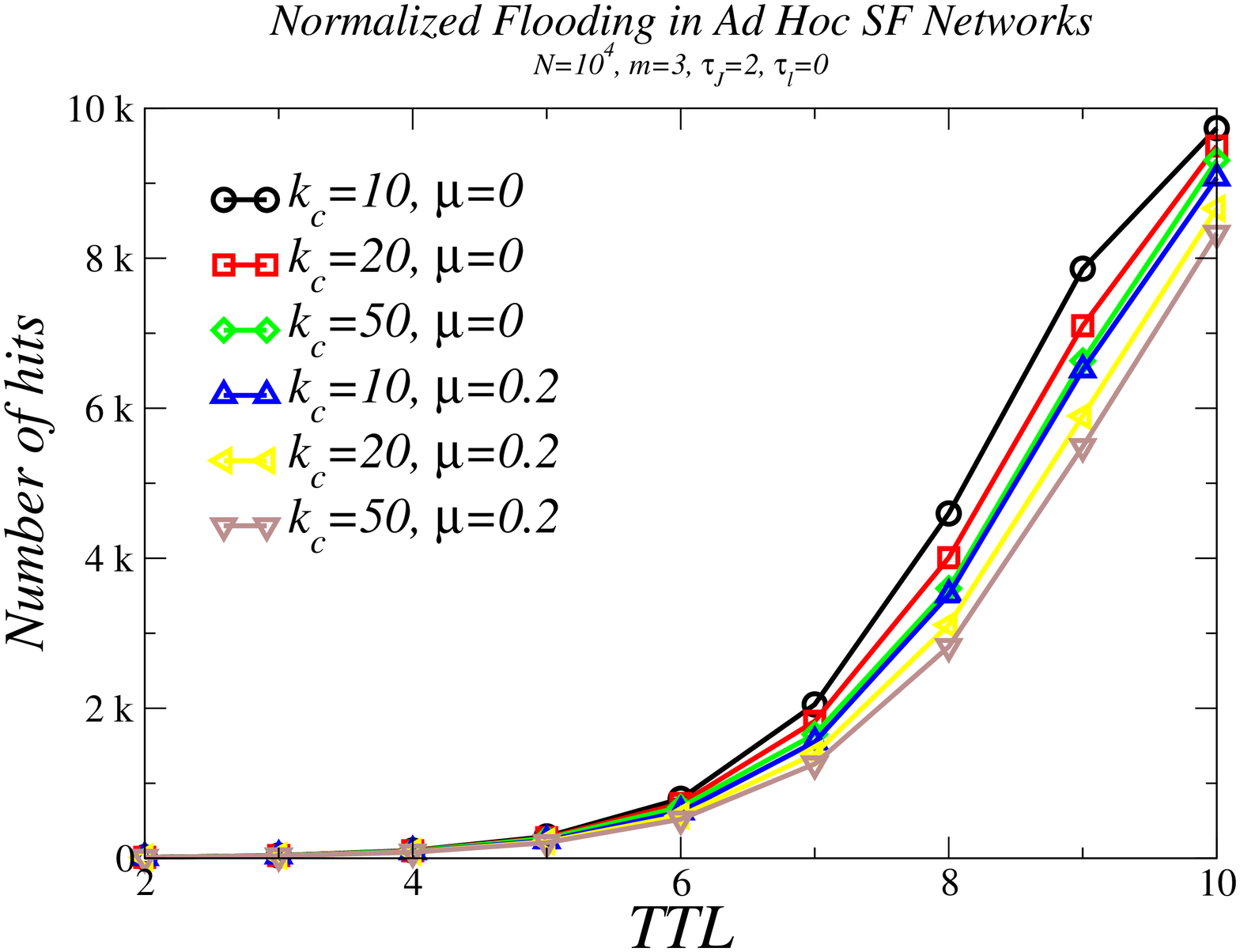} & \hspace{-5mm}
\includegraphics[keepaspectratio=true,angle=0,width=59mm]
{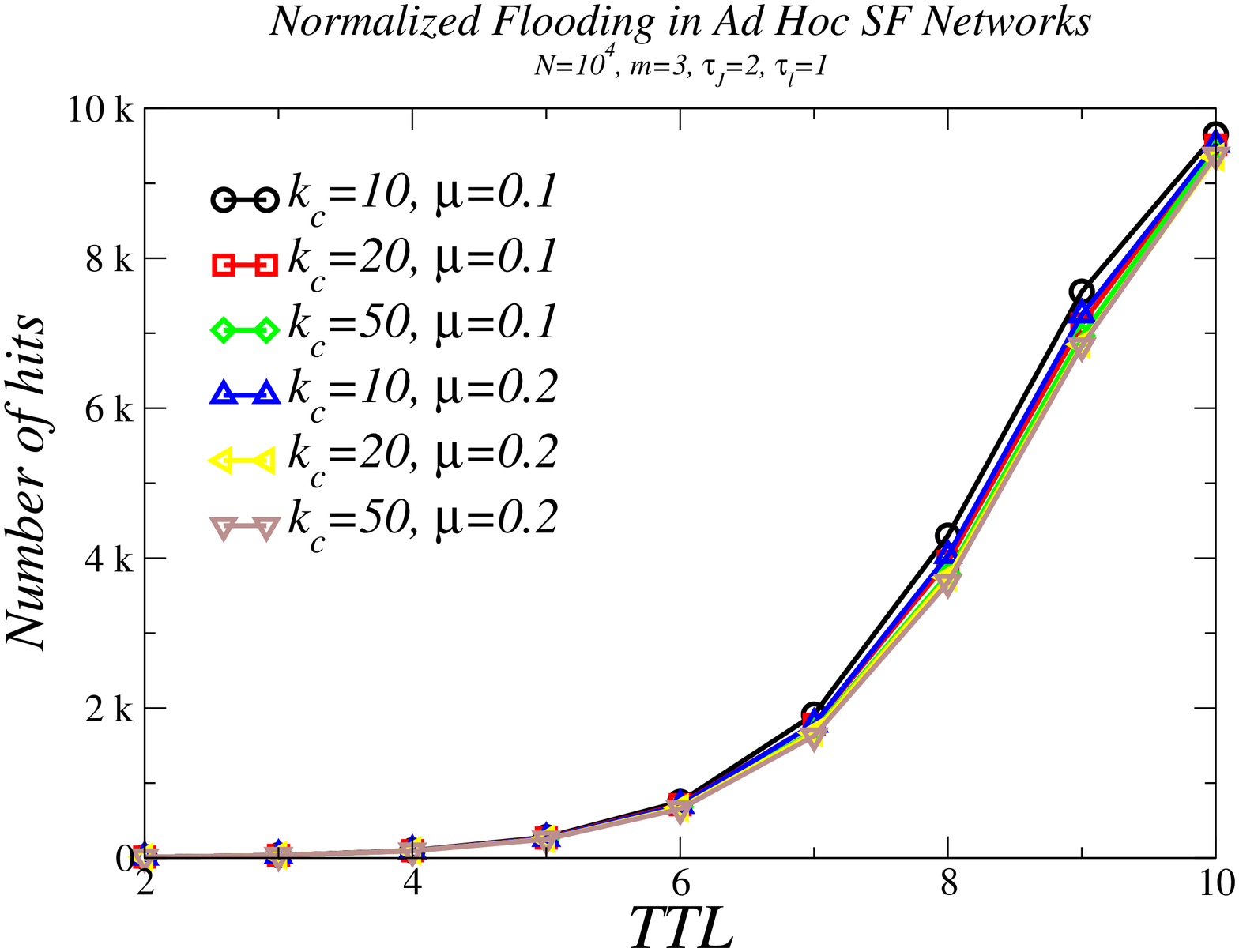} \vspace{-2mm} \\
\small{(a) $m$=1, $\tau_j$=2, $\tau_l$=2} & \small{(b) $m$=3,
$\tau_j$=2, $\tau_l$=0} &
\small{(c) $m$=3, $\tau_j$=2, $\tau_l$=1} \\
\end{tabular}
\end{center}
\vspace{-6mm} \caption{Normalized Flooding (NF) performance over
topologies generated with various $m$, $\tau_j$, and $\tau_l$
values.} \label{fig:NF} \vspace{-7mm}
\end{figure*}



%

\Section{Summary and Discussions} \label{sec:summary}

In summary, we worked on an ad-hoc limited scale-free network model
for unstructured peer-to-peer networks. We first developed localized
joining and leaving schemes for the peers and measure the efficiency
of search algorithms such as flooding and normalized flooding. By
considering the fact that the peers do want to store too many links
information we also imposed a hard cutoff on the degree a node can
have and analyzed its effect on the search efficiency. We
parameterized the locality of the joining and leaving schemes by two
parameters: $\tau_j$ (for joining) and $\tau_l$ (for leaving) which
are the number of hops nodes will use to construct sets of nodes
from which they will randomly choose other nodes and attempt to
connect by using the preferential attachment rules and by observing
the hard cutoff. Typically, high values of these parameters will
make the network a preferential attachment network with degree
distribution exponent $3$. We also modeled the random deletion of
the nodes by a probability parameter $\mu$.

Our search simulations show that the negative effects of the low
cutoff and high probability of deletion can be eased by increasing
the minimum degree in the network. This also helps one to avoid the
pathological case of $m$=1 for which the network will likely to have
isolated clusters hindering the efficiency of the search algorithms.
To remedy the negative effects of high values of $\mu$ which
destroys the scale-freeness in the network we enlarged the locality
of the leaving scheme, i.e., increasing $\tau_l$ for a fixed
$\tau_j$ and cutoff will increase the efficiency of normalized
flooding. Our findings are directly applicable to current
unstructured P2P networks in which the peers leave the network
unexpectedly and they have an upper limit for degree.

\section*{Acknowledgment} This work was supported by the U.S.
Department of Energy under contract DE-AC52-06NA25396 and by the
National Science Foundation under awards 0627039 and 0721542.
Authors would like to thank Sid Redner for fruitful discussions.


\bibliographystyle{latex8}
\bibliography{cutoff-adhoc}

\end{document}